\documentclass[12pt,twoside]{article}
	
	\usepackage{adjustbox}
	\usepackage{fancyhdr}
    \usepackage{url}
    \usepackage{hyperref}
    \usepackage{float}
	\usepackage[T1]{fontenc}
	\usepackage{graphicx}
	\usepackage[utf8]{inputenc}
	\usepackage{multicol}
	\usepackage{multirow}
    \usepackage{array}
	\usepackage[normalem]{ulem}
	\usepackage[svgnames]{xcolor}
    \usepackage{natbib}
    \usepackage{amsmath,amssymb}
    \usepackage[
    paperheight=11in,
    paperwidth=8.5in,
    left=20mm,
    right=20mm,
    top=20mm,
    bottom=25mm,
    footskip=1.2cm
    ]{geometry}
    
\begin{document}
 
	\vspace{2\baselineskip}
 
	\begin{figure}[H]
	\includegraphics[width=4.4cm,height=2.81cm]{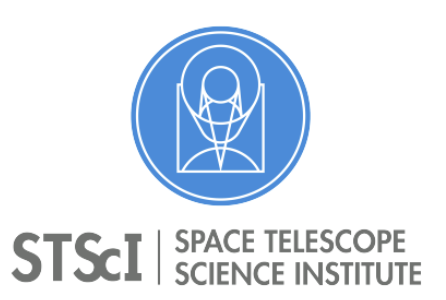}
	\end{figure}

	\vspace{1\baselineskip}
 
    \begin{center}
        {\LARGE \textbf{JWST TECHNICAL REPORT}}
    \end{center}
	
	\begin{table}[H]
    {
    \setlength\arrayrulewidth{1.5pt}
	\begin{tabular}{|m{0.6 truein} m{2.5 truein} | m{0.9 truein} m{2.2 truein}|}
	\hline
	Title: & Systematic Slopes in NIRSpec BOTS Observations  & Doc: & JWST-STScI-009489, SM12$\#$\\
      &  & Date:  & 03 September 2026 \\
      & & Rev: & - \\ 
	\hline
	Authors: & Munazza K. Alam, Leonardo Ubeda, Sophie von Coelln, Ceiligh Cacho-Nagrete, Nestor Espinoza & Release~Date: & 09 October 2026 \\ 
             & & & \\
             &  & &  \\
    Phone: & +1 410-338-4445 &  &  \\
	\hline
	\end{tabular}
    }
	\end{table}

	\vspace{1\baselineskip}
 
	\section*{Abstract}
    
   We investigate empirical white light and spectroscopic trends in the temporal systematic slopes seen in NIRSpec Bright Object Time Series (BOTS) observations. Using a sample of 14 high-resolution G395H (dispersed across both the NRS1 and NRS2 detectors) and 10 low-resolution Prism (dispersed across NRS1 only) transit light curves, we measured the visit-long, pre-transit, and post-transit slopes. For the G395H white light curves, we find a correlation between the G395H NRS1 systematic slope and the host star magnitude, as well as the number of groups per integration. We also note that the systematic slopes vary with time over the course of a given time-series, with steeper pre-transit slopes compared to post-transit. These trends in the white light slopes are not apparent for G395H NRS2 or Prism. At the spectroscopic light curve level, the NRS1 and Prism slopes show a distinct wavelength-dependent signature across the full sample with the largest slopes between $\sim$3--3.4 $\mu$m. At these wavelengths, the brightest targets display the strongest chromatic slope signatures. The NRS2 slopes show no apparent chromaticity across the full sample. 
 
	\vspace{1\baselineskip}

\section{Introduction \label{sec:Introduction}}
\vspace{-2mm}

NIRSpec Bright Object Time Series (BOTS) is the most popular JWST instrument mode for observations of transiting exoplanets ranging from gas giants to rocky planets, thanks to its exquisite precision, resolution, and wavelength coverage of critical molecular species (e.g., CO$_{2}$, H$_{2}$O, CH$_{4}$). Interpretations of NIRSpec transit spectra have displayed instrumental systematics such as offsets between the NRS1 and NRS2 detectors for high-resolution observations \citep[e.g.,][]{Damiano24}, lower-than-expected transit-depth precisions \citep[e.g.,][]{Alam25}, and visit-long systematic slopes \citep[e.g.,][]{Luque25}. 

Such visit-long systematic slopes were first observed in NIRSpec BOTS observations during JWST commissioning \citep{Espinoza23}, and have been consistently observed since then throughout the first few JWST GO cycles -- with varying slope strengths seen across datasets. While these slopes are typically accounted for in differential transit depth measurements \citep[e.g.,][]{Espinoza23,Damiano24,Alam25}, the underlying instrumental systematic effect causing these slopes is not fully understood. 
Qualitatively, the slope appears more strongly on the NRS1 detector in observations taken with the high-resolution gratings (e.g., G395H, G235H) as well as with the low-resolution Prism filter (which is dispersed on NRS1 only) compared to the NRS2 detector. The slope at the red end of NRS1 also appears to be different from the slope at the blue end of NRS2, suggesting a wavelength-dependence in the strength of these slopes. 

Initial work was performed internally at STScI to correlate the white light and spectroscopic systematic slopes for a single BOTS dataset against engineering parameters, but no significant correlations were found. The repeatability of this slope also remains unknown, as multiple observations of the same target may yield different measured visit-long slopes. 
Furthermore, partial phase curves -- which require flux stability over long timescales -- are of interest to the community \citep[e.g.,][]{LustigYaeger21} and would require better calibration of the baseline flux for NIRSpec BOTS data. However, such observations would be impacted by the presence of visit-long systematic slopes. As such, further investigation on these slopes is critical for enhancing atmospheric inferences derived from BOTS observations. 

In this Technical Report, we measure the white light (i.e., band-integrated) and spectroscopic slopes for a sample of 14 G395H and 10 Prism transit observations to investigate how these slopes might vary due to target properties (e.g., host star brightness), and/or observatory/detector read properties (e.g., observation date, groups per integration, number of integrations). An outline of this report is as follows. In Section \ref{sec:analysis}, we describe the sample selection, data reduction, and light curve fitting methods. The results from the white light and spectroscopic slopes analyses for G395H and Prism are detailed in Section \ref{sec:results}. We contextualize and summarize this study in Section \ref{sec:disc_summary}.

\section{Analysis}
\label{sec:analysis}

\subsection{Sample Selection}
We selected a sample of planets with published NIRSpec BOTS observations that span a range of host star spectral types and planetary parameters. As shown in Table \ref{tab:sample}, this sample includes 14 G395H and 10 Prism observations for targets ranging from giant planets to super-Earths.

\begin{table}[ht]
\centering
\begin{tabular}{c c c c c c}
\hline
Planet & Host Star & J mag & PID & NRS1 Slope & NRS2 Slope \\
Name & Spectral Type & & & (ppm\,hr$^{-1}$) & (ppm\,hr$^{-1}$) \\
\hline
\multicolumn{4}{l}{\textit{Prism}} \\
\hline
WASP-39b   & G8 & 10.66 & 1366 & $-$130 & --- \\
WASP-6b    & G8 & 10.77 & 5924 & $-$143 & --- \\
WASP-19b   & G8 & 10.91 & 5924 & $-$64 & --- \\
TOI-1899b  & M0 & 11.34 & 4227 & $-$204 & --- \\
Kepler-86b & F9 & 11.5 & 3235  & $-$32 & --- \\ 
TOI-3984b  & M4 & 11.93 & 3171 & $-$101 & --- \\
TOI-5205b  & M4 & 11.93 & 3171 & $-$170 & --- \\
HATS-6b    & M1 & 12.04 & 3171 & $-$126 & --- \\
LP890-9b   & M6 & 12.26 & 7073 & $-$110 & --- \\
HATS-75b   & M0 & 12.48 & 3171 & $-$117 & --- \\

\hline
\multicolumn{4}{l}{\textit{G395H}} \\
\hline
GJ486b   & M3.5 & 7.20 & 1981  & $-$337 & $-$10 \\
GJ357b   & M2.5 & 7.34 & 2512  &$-$287 & $-$36 \\
TOI-836b & K7   & 7.58 & 2512  &$-$269 & 12 \\
TOI-836c & K7   & 7.58 & 2512  &$-$205 & 12 \\
L168-9b  & M1   & 7.94 & 2512  &$-$175 & 19 \\
TOI-776b & M1   & 8.48 & 2512  &$-$103 & 11 \\
TOI-776c & M1   & 8.48 & 2512  &$-$127 & 0 \\
HAT-P-1b & G0   & 9.16 & 5924  &$-$61 & 38 \\
GJ1132b  & M4   & 9.25 & 1981  &$-$64 & $-$19 \\
WASP-107b & K7  & 9.38 & 1201  &$-$54 & $-$13 \\
WASP-178b & A1  & 9.78 & 2055  &$-$52 & 1 \\
WASP-63b & G8   & 9.82 & 5924  &$-$13 & 7 \\
HAT-P-26b & K1  & 10.08 & 1312 &$-$20 & 3  \\
WASP-17b & F6   & 10.51 & 1353 &4 & 18  \\

\hline
\end{tabular}
\caption{Sample of targets with published NIRSpec BOTS Prism and G395H observations used in the systematic slopes analysis.}
\label{tab:sample}
\end{table}

\subsection{Data Reduction} 

Each G395H and Prism transit dataset was reduced using the open-source python-based {\tt ExoTiC-JEDI} pipeline \citep{Alderson22} to produce the 1D time-series stellar spectra as well as the NRS1 (for Prism and G395H) and NRS2 (for G395H only) white light curves. In the G395H analysis, the detectors were treated as separate visits. 

Using a modified version of Stage 1 of the {\tt jwst} pipeline (version 1.18.0, context map 1364); \citealt{Bushouse22}), we performed the standard corrections for linearity, dark current, and saturation, and applied a jump detection threshold of 15. We next performed a custom group-level $1/f$ correction, masking the spectral trace 15$\sigma$ from the dispersion axis for each integration and subtracting the median pixel value of non-masked pixels from each detector column in each group. We also applied the {\tt ExoTiC-JEDI} custom bias subtraction before the standard ramp fitting step. To track the $x$- and $y$-pixel positional shifts on the detector, we cross-correlated the 1D time-series stellar spectra. We extracted the wavelength solution using Stage 2 of the {\tt jwst} pipeline to produce the 2D wavelength map. We generated the broadband (white light) light curves by summing up the stellar flux in the time-series across the full wavelength range for G395H (NRS1, 2.814–3.717 $\mu$m; NRS2, 3.824–5.111 $\mu$m) and Prism (0.6--5.3\,$\mu$m). For the G395H spectroscopic light curves, we generated 54 binned light curves with uniform widths of 60 pixels ($\sim$0.2 $\mu$m wide) based on the G395H wavelength solution. For the Prism spectroscopic light curves, we binned the light curves into 116 spectroscopic channels of varying widths (0.01 $\mu$m wide in the blue and 0.08 $\mu$m wide in the red) with roughly equal counts in each. 



\subsection{Light Curve Fitting}

We fit the broadband and spectroscopic light curves for NRS1 and NRS2 following the methods of \citet{Alam25}. We fit each light curve with a two-component function consisting of an analytic transit model generated using {\tt batman} \citep{Kreidberg15} multiplied by a systematics model of the form:

\begin{equation}
    S(\lambda) =  c_{0} + (c_{1} \times x |y|) + (c_{2} \times t)
\end{equation}

\noindent where  $x$ and $y$ are the $x$- and $y$- pixel position shifts on the detector, $t$ is time, and  the coefficients $c_{0}$, $c_{1}$, and $c_{2}$ are constants to be fit. Here, $c_{2}$ is the visit-long slope. 

For the white light curves, we fixed the period $P$ to literature values, set the eccentricity $e$ to zero, and argument of periastron $\omega$ to 90, and fit for the mid-transit time $T_{0}$, the scaled semi-major axis $a/R_{\star}$, inclination $i$, planet-to-star radius ratio $R_{p}/R_{\star}$, and stellar baseline flux using wide uniform priors. We held the nonlinear limb-darkening coefficients \citet{Claret00} fixed to theoretical values, which we computed using {\tt ExoTiC-LD} \citep{Grant22} with Set One of the MPS-ATLAS stellar models \citet{Kostogryz22,Kostogryz23} and the stellar effective temperature $\rm T_{eff}$, surface gravity log(g), and metallicity.  For the spectroscopic light curves, we fixed $T_{0}$, $a/R_{\star}$, and $i$ to the best-fit values from the broadband light curves and fit for $R_{p}/R_{\star}$. We adopted the optimized parameters and their standard deviations as the best-fitting value and associated uncertainty for our fitted parameters. For each fitted light curve, we extracted the visit-long systematic slope $c_{2}$ and converted to units of ppm hr$^{-1}$ for this analysis. We also measured the pre-transit and post-transit slopes by fitting a line to the out-of-transit baseline pre- and post-transit, respectively, and converting to ppm hr$^{-1}$. 



\begin{figure}[!ht]
    \centering
    \includegraphics[width=\linewidth]{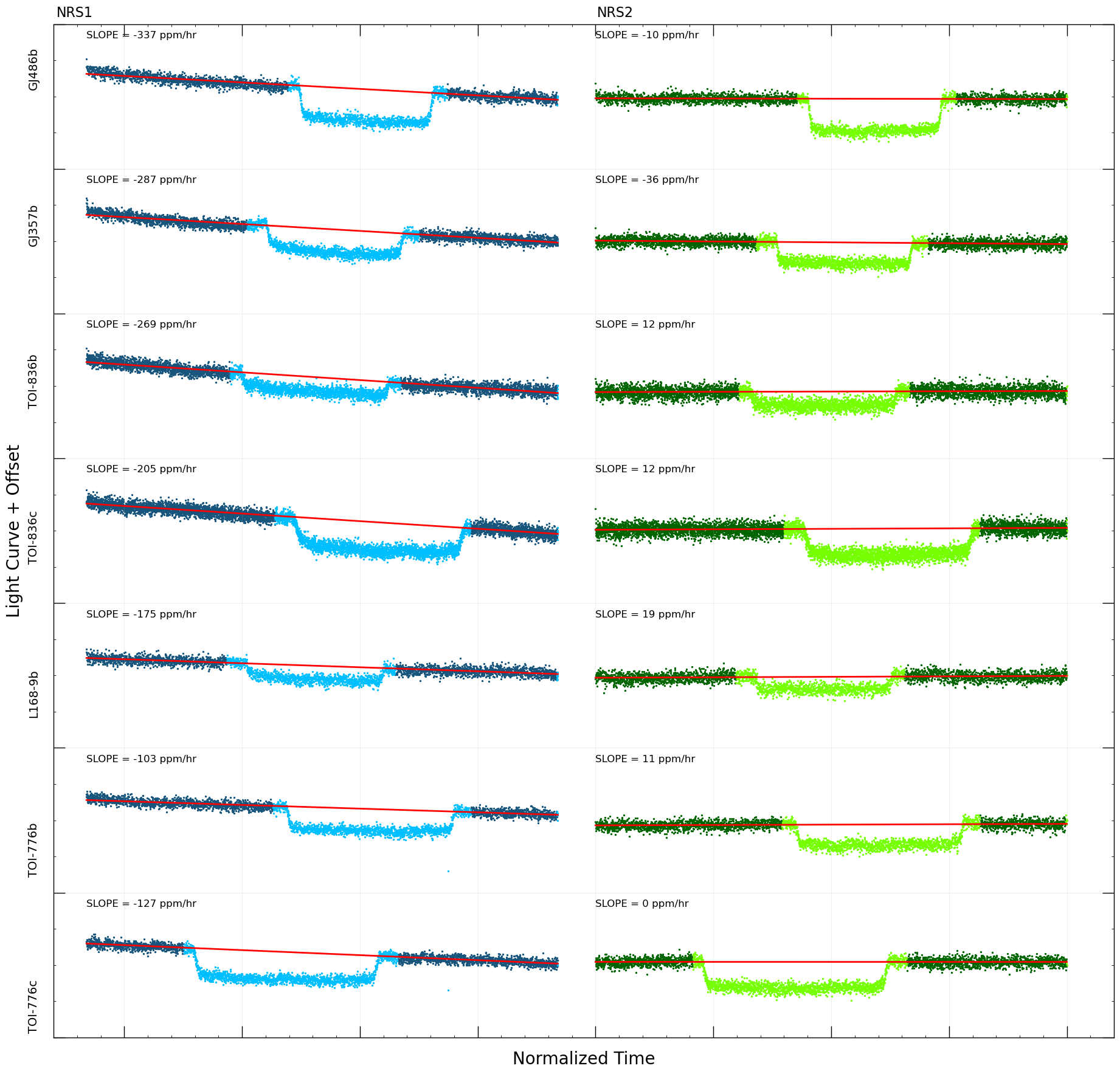}
    \caption{NRS1 (left; blue) and NRS2 (right; light green) white light curves for the brightest (J $<$ 9) targets in the G395H sample. The fitted visit-long out-of-transit linear trend is shown in red, with the measured visit-long slopes annotated.}  
    \label{fig:g395h_spec_slopes_bright}
\end{figure} 

\begin{figure}[!ht]
    \centering
    \includegraphics[width=\linewidth]{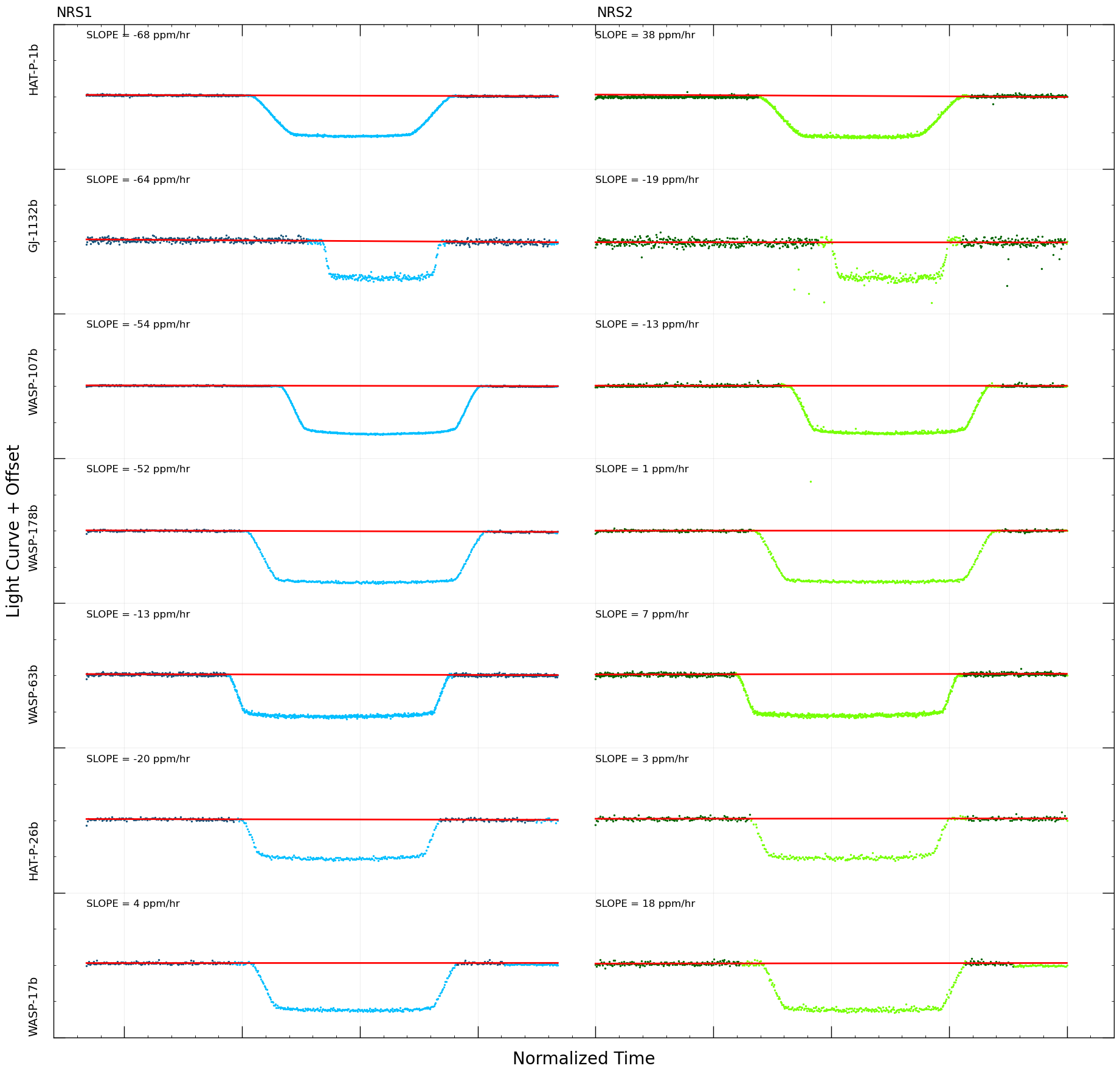}
    \caption{Same as Figure \ref{fig:g395h_spec_slopes_bright}, but for the fainter (J $>$ 9) targets in the G395H sample.}  
    \label{fig:g395h_spec_slopes_fainter}
\end{figure} 

\begin{figure}[!ht]
    \centering
    \includegraphics[width=\linewidth]{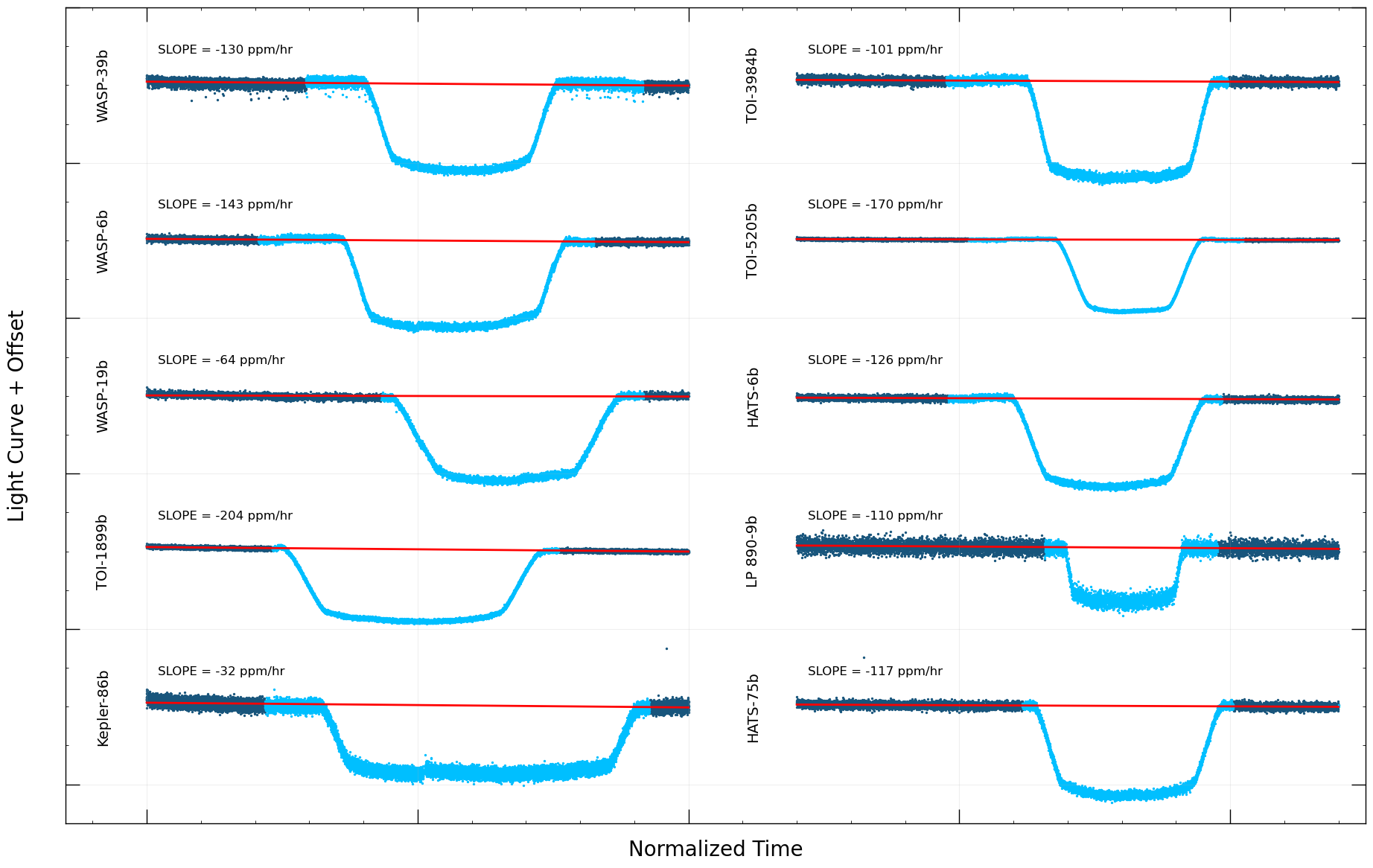}
    \caption{White light curves (blue) for the Prism sample used in our analysis. The fitted visit-long out-of-transit linear trend is shown in red, and the measured slopes from these fits are annotated.}  
    \label{fig:prism_light_curves}
\end{figure}

\section{Results}
\label{sec:results}

The reduced white light curves and fitted linear trends are shown in Figures \ref{fig:g395h_spec_slopes_bright} and \ref{fig:g395h_spec_slopes_fainter} for the G395H sample, and in Figure \ref{fig:prism_light_curves} for the Prism sample. The measured visit-long systematic slopes for the white light curves are given in Table \ref{tab:sample}. As is visually apparent, the visit-long systematic slopes for G395H NRS1 span a broad range (4 to $-$337 ppm hr$^{-1}$) across the full sample. The Prism NRS1 slopes appear to be similar in strength ($-32$ to $-$204 ppm hr$^{-1}$) to the G395H NRS1 slopes, but do not span as wide a range. The G395H NRS2 slopes are weaker than those measured for G395H NRS1, spanning only $\sim \pm$ 40 ppm hr$^{-1}$. 

We also searched for empirical trends in the measured visit-long slopes at both the white light and spectroscopic light curve levels for several parameters of interest, including the number of groups per integration, magnitude of the host star, number of integrations (as a proxy for observation duration), stellar effective temperature (as a proxy for host star spectral type), planetary radius, and date of observation. Our results for the white light slopes analysis are described in Section \ref{sec:wl}, and the spectroscopic slope results are detailed in Section \ref{sec:spec}. 

\subsection{White Light Slopes}
\label{sec:wl}

\subsubsection{G395H}
 
\begin{figure}[!ht]
    \centering
    \includegraphics[width=\linewidth]{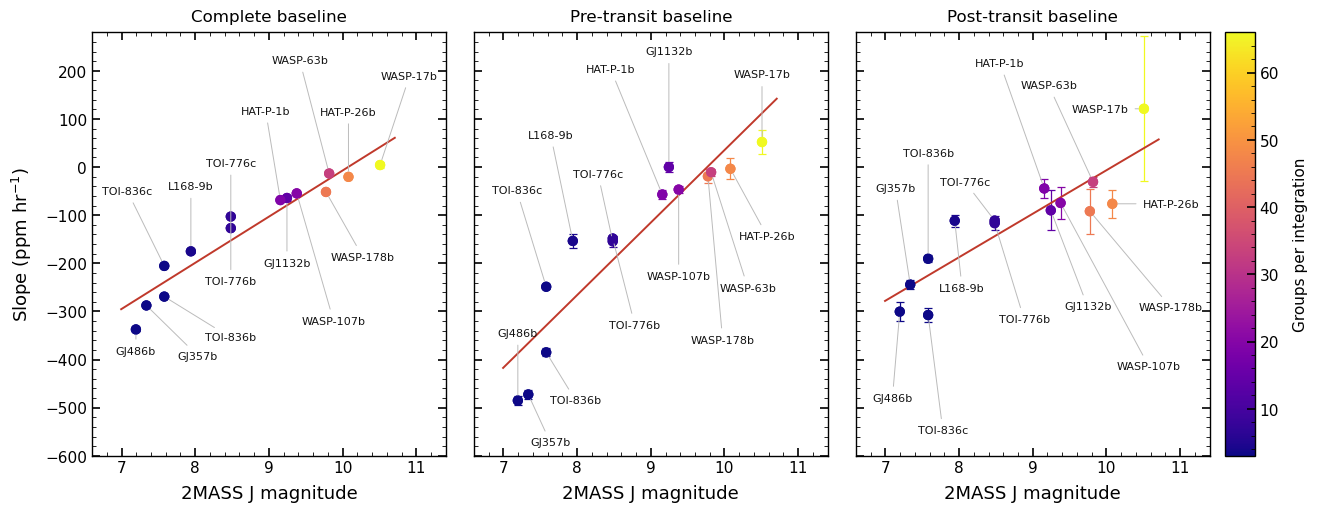}
    \caption{The G395H NRS1 visit-long (left), pre-transit (middle), and post-transit (right) systematic slopes as a function of host star J magnitude for a sample of 14 planets (points) color coded by number of groups per integration. Brighter targets with observations using fewer groups per integration reveal stronger visit-long systematic slopes in the time series observations.}
    \label{fig:nrs1_wlc_slopes}
\end{figure}

For the G395H white light curve slopes, we found a correlation between the G395H NRS1 systematic slope and the host star 2MASS J magnitude, as well as the G395H NRS1 systematic slope and the number of groups per integration. These correlations are linked, since brighter targets typically require less up-the-ramp groups per integration. As shown in Figure \ref{fig:nrs1_wlc_slopes}, brighter targets appear to have stronger G395H NRS1 slopes -- as do targets with fewer groups per integration. No other empirical correlations were found for the other target and detector properties that we investigated. In addition to the visit-long slope, we also measured the pre- and post-transit slopes for each transit light curve to explore how the systematic slopes vary over the course of the time-series observations. We found that the systematic slopes vary with time over the course of a given transit observation, with steeper pre-transit slopes compared to post-transit (Figure \ref{fig:nrs1_wlc_slopes}).   

\begin{figure}[!ht]
    \centering
    \includegraphics[width=\linewidth]{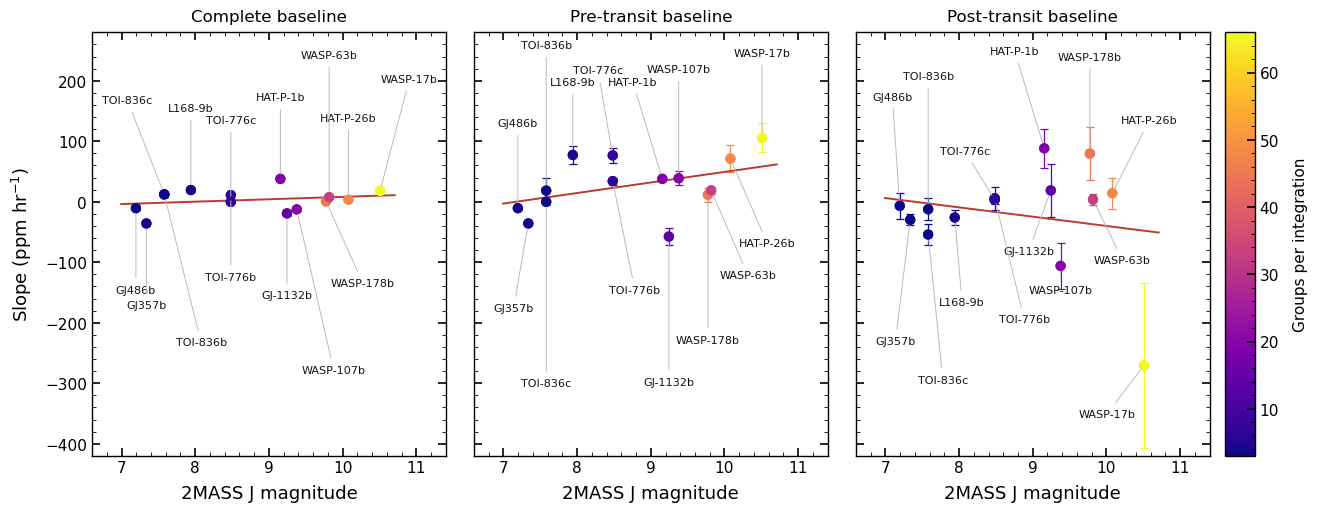}
    \caption{The G395H NRS2 visit-long (left), pre-transit (middle), and post-transit (right) systematic slopes as a function of host star J magnitude for a sample of 14 planets (points) color coded by number of groups per integration. The NRS2 slopes do not display any clear trends with target or observatory/detector read properties.}
    \label{fig:nrs2_wlc_slopes}
\end{figure}

Notably, these trends in the white light slopes are not apparent for G395H NRS2. As shown in Figure \ref{fig:nrs2_wlc_slopes}, the G395H NRS2 visit-long slopes are weak ($\sim \pm$40 ppm hr$^{-1}$) and do not vary widely across the full sample, whereas the pre- and post-transit slopes show significant scatter. Within this scatter, the pre-transit G395H NRS2 slopes show a weak correlation with magnitude and groups per integration. The post-transit slopes, however, show significant scatter with WASP-17b as an outlier -- although the post-transit baseline in our reduction for this target shows a flux offset that is likely driving the measured slope value and its large uncertainty. 

\subsubsection{Prism}

\begin{figure}[!ht]
    \centering
    \includegraphics[width=\linewidth]{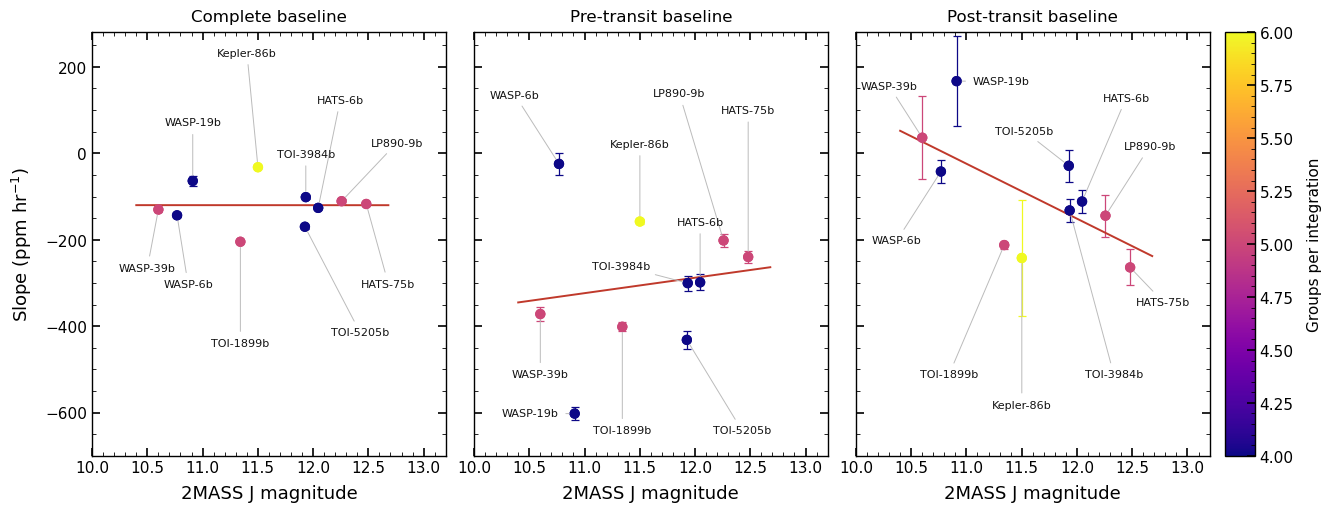}
    \caption{The Prism visit-long (left), pre-transit (middle), and post-transit (right) systematic slopes as a function of J magnitude for a sample of 10 planets (points) color coded by number of groups per integration. The systematic slopes across the full sample do not show strong correlations with host star magnitude or number of groups per integration. } 
    \label{fig:prism_wlc_slopes}
\end{figure}

We also investigated the broadband slopes for our sample of Prism observations (see Table \ref{tab:sample}). Figure \ref{fig:prism_wlc_slopes} shows the measured systematic slopes for the Prism sample, which demonstrate that all of the targets in our sample display negative visit-long slopes similar in magnitude to our G395H NRS1 measured slopes. However, we do not find any strong empirical correlations across the full sample with number of groups per integration, host star magnitude, or any of the other target and detector properties we explored. 

\subsection{Spectroscopic Slopes}
\label{sec:spec}

\subsubsection{G395H (NRS1 \& NRS2)}

\begin{figure}[!ht]
    \centering
    \includegraphics[width=\linewidth]{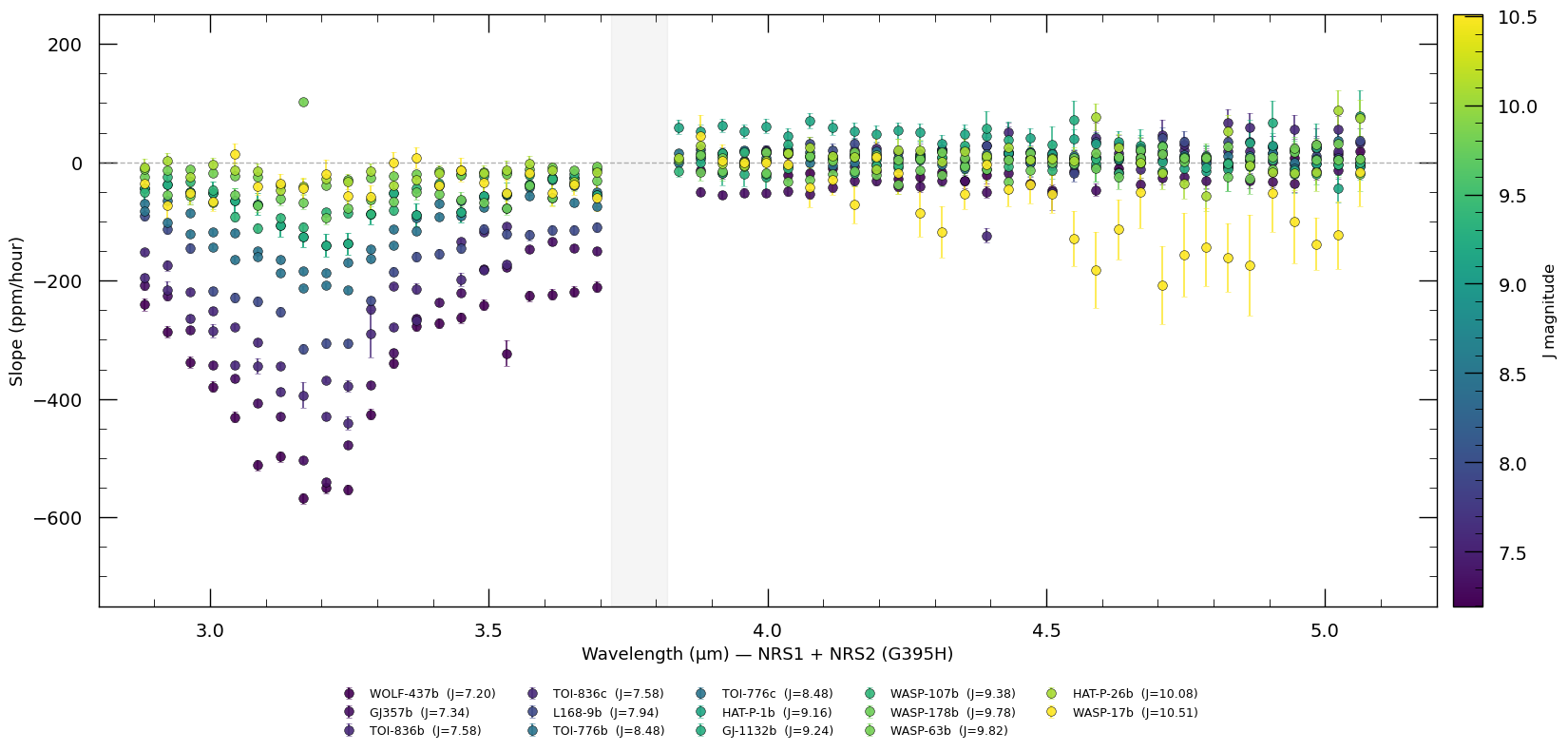}
    \includegraphics[width=\linewidth]{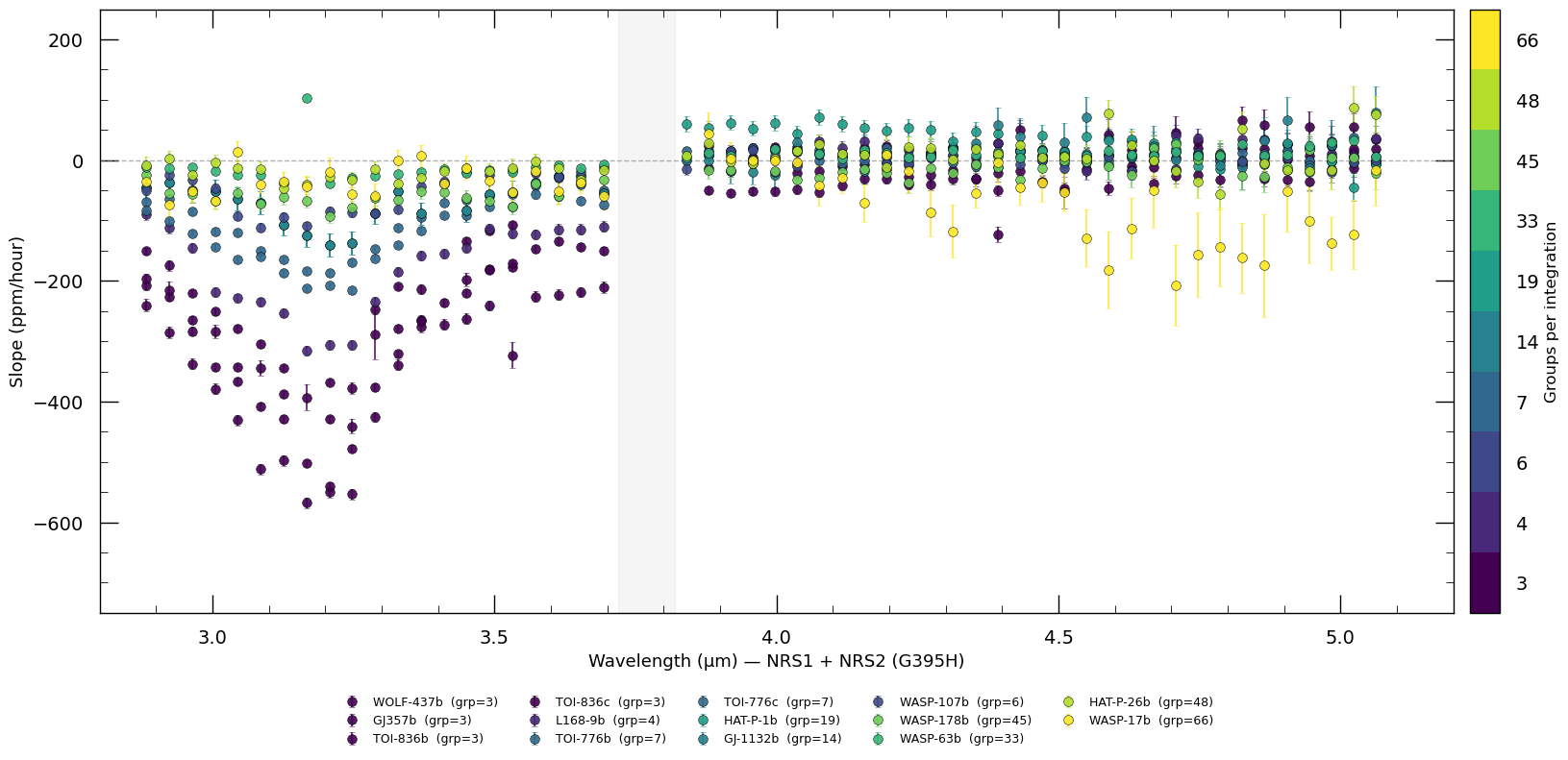}
    \caption{The visit-long G395H spectroscopic slopes for NRS1 and NRS2 color coded by J magnitude (top) and number of groups per integration (bottom), with the detector wavelength gap shown in the gray shaded region. The G395H spectroscopic slopes reveal a distinct chromatic signature across the full sample with the largest slopes between $\sim$3--3.4 $\mu$m. At these wavelengths, the brightest targets display the strongest chromatic slope signatures. The NRS2 slopes show no apparent chromaticity in slopes for all targets. 
    }
    \label{fig:g395h_spec_slopes}
\end{figure}

We then investigated the systematic slopes at the spectroscopic light curve level to explore how slope strength varies across wavelength. We measured the visit-long spectroscopic light curve slopes for each detector, and color coded the targets by the J magnitude of the host star as well as the number of groups per integration which -- based on our findings from Section \ref{sec:wl} -- have been shown at the white light curve level to be strongly correlated with the strength of the visit-long slope in the time-series). These results are shown in Figure \ref{fig:g395h_spec_slopes}.  

The NRS1 and NRS2 detectors reveal different wavelength-dependent (chromatic) systematic slopes.  In particular, the G395H NRS1 slopes show a distinct chromatic signature between 2.8--3.7\,$\mu$m across all of the targets with strongest slopes between $\sim$3-3.4 $\mu$m (with a peak at $\sim$3.2 $\mu$m). The G395H NRS2 slopes do not show evidence of chromaticity for our full sample. For this detector, the slopes are $\sim$0 ppm hr$^{-1}$ at all wavelengths, i.e., we do not find any apparent chromaticity in the spectroscopic slopes across the full NRS2 wavelength range of 3.8--5.1\,$\mu$m. 

\subsubsection{Prism (NRS1 only)}

\begin{figure}[!ht]
    \centering
    \includegraphics[width=\linewidth]{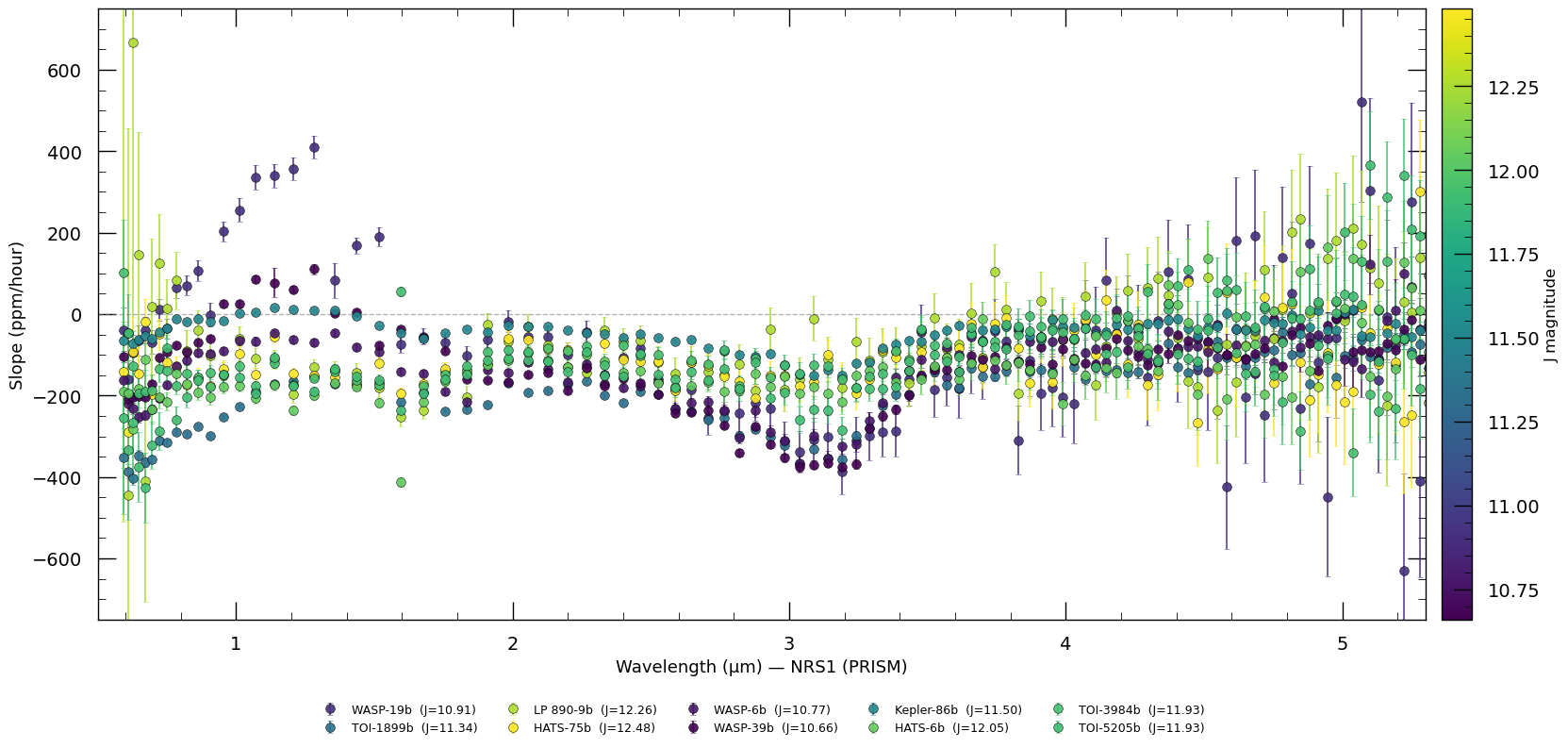}
    \includegraphics[width=\linewidth]{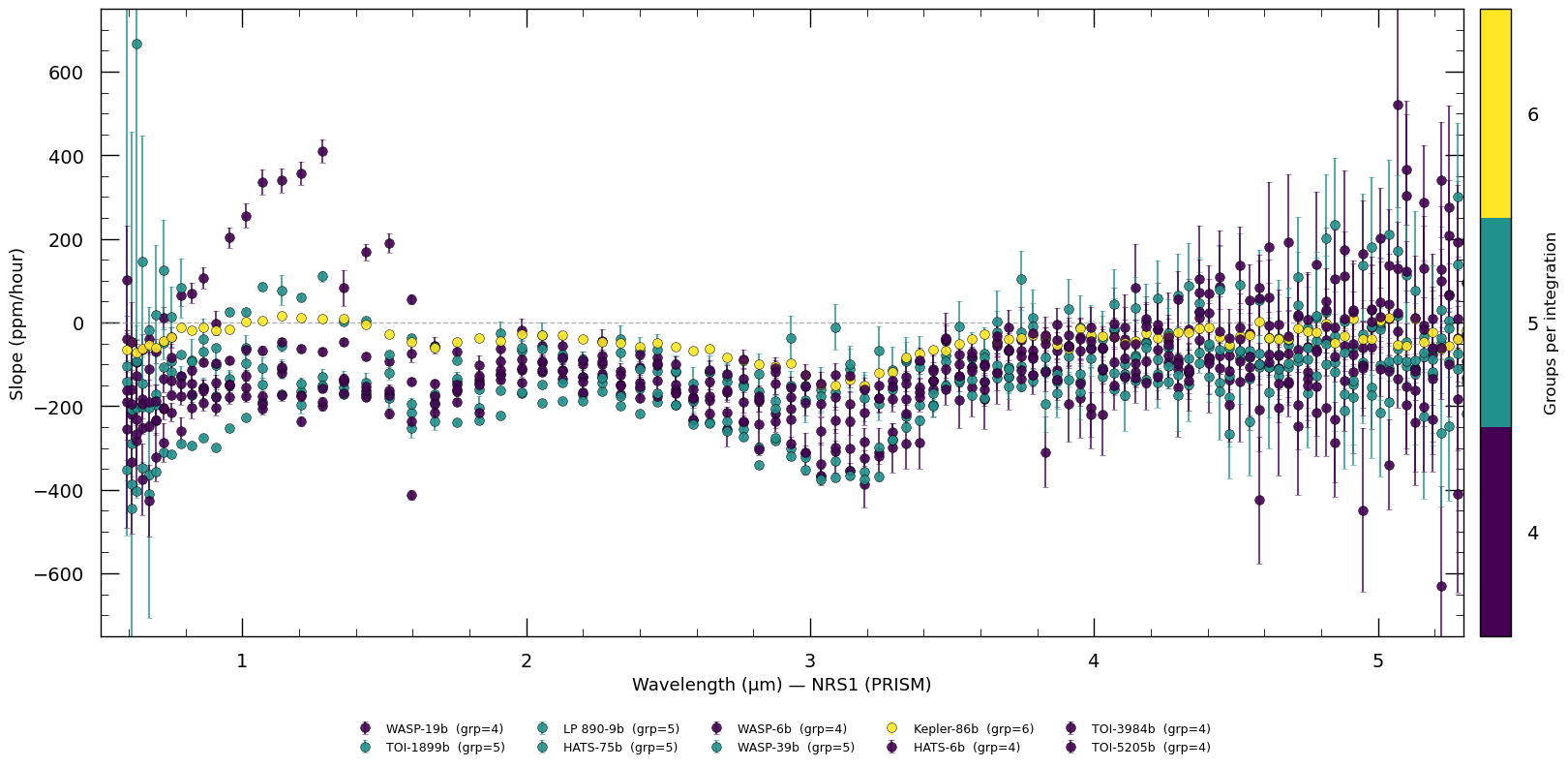}
    \caption{The visit-long spectroscopic slopes for the Prism sample color coded by J magnitude (top) and number of groups per integration (bottom). We note the same distinct chromatic signature between $\sim$3--3.4 $\mu$m seen in the G395H NRS1 spectroscopic slopes, as well as another chromatic signature between 1.4--1.8\,$\mu$m.  }
    \label{fig:prism_spec_slopes}
\end{figure}

We also measured the spectroscopic slopes for our Prism sample, which are shown in Figure \ref{fig:prism_spec_slopes}. Prism spans a wide wavelength range (0.6--5.3\,$\mu$m) and is dispersed on NRS1 only -- as are the medium-resolution (e.g., G235M, G395M) NIRSpec gratings. Intriguingly, we find the same distinct chromatic signature seen for G395H NRS1 of stronger spectroscopic slopes between $\sim$3--3.4 $\mu$m and a peak in the slope strengths in this wavelength range at $\sim$3.2\,$\mu$m. Additionally, we note another chromatic signature at bluer wavelengths which is characterized by stronger slopes between 1.4--1.8\,$\mu$m and a peak at 1.6\,$\mu$m.

\section{Discussion \& Summary}
\label{sec:disc_summary}

These results mark the first empirical evidence of higher levels of instrumental systematics in BOTS observations of brighter targets across a sample of NIRSpec transit data. The correlation between host star magnitude and the strength of the visit-long systematic slopes for G395H NRS1 (Figure \ref{fig:nrs1_wlc_slopes}) presents compelling evidence that time-series observations of brighter targets -- which require a fewer number of groups per integration -- are more likely to have stronger visit-long slopes in the transit light curves compared to fainter targets. These brighter targets host small planets as well, consistent with published findings that observations of small planets around bright stars show stronger G395H NRS1 detector-level systematics \citep[e.g.,][]{May2023,Alam25,Alderson2025}. 
The time-dependence of the slopes, as demonstrated by the stronger pre-transit slopes compared to those post-transit (Figures \ref{fig:nrs1_wlc_slopes}, \ref{fig:nrs2_wlc_slopes}, and \ref{fig:prism_wlc_slopes}), suggest that this systematic may be related to detector settling effects, which are the strongest at the beginning of the time-series observation. 

The G395H NRS1 spectroscopic slopes appear to be steeper at wavelengths where the observed count rate is the highest. However, since the distinct chromatic slope signatures appear for both G395H NRS1 as well as for Prism (NRS1 only) in the same wavelength range, this chromaticity may be an NRS1-specific instrumental systematic rather than a function of the observed count rate. Alternatively, the chromaticity of the systematic slopes may be linked to the observed count rate of a given target/visit, i.e., different targets with similar count rates may produce similar spectroscopic slopes. In this case, the white light slope trends seen in Figures \ref{fig:nrs1_wlc_slopes}, \ref{fig:nrs2_wlc_slopes}, and \ref{fig:prism_wlc_slopes} would map to the spectroscopic trends shown in Figures \ref{fig:g395h_spec_slopes} and \ref{fig:prism_spec_slopes}. It is also possible that the slopes may potentially be linked to the detector's exposure history, as NRS1 is exposed more often (across Prism as well as the medium- and high-resolution gratings) compared to NRS2 (which is only exposed during H-grating observations).  

Furthermore, the presence of visit-long systematic slopes for NRS1, in tandem with the lack of such slopes across the full sample for NRS2, may be related to the fact that the two detectors -- despite being otherwise identical in their construction -- were reported in pre-launch ground-testing to have different electron trap densities, with higher trap densities reported for NRS1 compared to NRS2 (priv.~comm). These trap density differences between the detectors may provide an explanation for the appearance of the systematic slopes on NRS1 across Prism and the higher-resolution gratings, but not in NRS2 for observations taken with the high-resolution gratings. Furthermore, while the data that we have analyzed here can be well-fit with a linear slope, this sample is not sufficient to rule out if there is a non-linear component to the visit-long systematic slopes seen in BOTS observations. 

Observations of brighter objects typically utilize the G395H grating, since these targets are often brighter than the Prism bright limit of J$\sim$11 for the nominal SUB512 subarray. With the new BOTS Prism multistripe subarrays\footnote{\href{https://jwst-docs.stsci.edu/jwst-near-infrared-spectrograph/nirspec-instrumentation/nirspec-detectors/nirspec-detector-readout\#NIRSpecDetectorReadout-bots_multistripeNIRSpecBOTSmultistripedetectorreadout}{NIRSpec BOTS multistripe detector readout}} that will be available beginning in JWST Cycle 6, the 3-group bright limit for Prism has been pushed down to J$\sim$8.3 for the smallest (SUB32$\textunderscore$M16$\textunderscore$PRISM) multistripe subarray. Future Prism multistripe observations may therefore display stronger visit-long systematic slopes in the transit light curves than nominal Prism BOTS observations.

\bibliography{references}

\end{document}